# Giant Magnetic Fluctuations at the Critical Endpoint in Insulating $HoMnO_3$

Y. J. Choi[1,2], N. Lee[1,2], P. A. Sharma[1†], S. B. Kim[1§], O. P. Vajk[3], J. W. Lynn[3], Y. S. Oh[1], and S-W. Cheong[1]

[1]*Rutgers Center for Emergent Materials and Department of Physics & Astronomy, Rutgers University, Piscataway, New Jersey, 08854, USA*

[2]*Department of Physics and IPAP, Yonsei University, Seoul 120-749, Korea*

[3]*NIST Center for Neutron Research, National Institute of Standards and Technology, Gaithersburg, Maryland 20899, USA*

Although abundant research has focused recently on the quantum criticality of itinerant magnets, critical phenomena of insulating magnets in the vicinity of critical endpoints (CEP's) have rarely been revealed. Here we observe an emergent CEP at 2.05 T and 2.2 K with a suppressed thermal conductivity and concomitant strong critical fluctuations evident via a divergent magnetic susceptibility (e.g., $\chi''$(2.05 T, 2.2 K)/$\chi''$(3 T, 2.2 K)≈23,500 %, comparable to the critical opalescence in water) in the hexagonal insulating antiferromagnet $HoMnO_3$.

[†]Present address: Materials Physics Department, Sandia National Laboratories, Albuquerque, NM 87123, USA

[§]Present address: Advancement for College Education Center, Konyang University, Chungnam 320-711, Korea

One of the fundamental processes in condensed matter systems concerns transitions from one state of matter to another[1, 2]. A first order transition which does not involve a symmetry change between two phases can give rise to a transition line in the phase diagram that terminates at a CEP[3, 4]. The ideas behind phase boundaries and CEP's naturally play an essential role in the descriptions of a wide range of phenomena ranging from quantum chromodynamics[5-7] to cell death in anticancer therapies[8]. A familiar example occurs for the transition between the liquid and vapor phases of water, which takes place without a change of symmetry and where strong critical fluctuations known as critical opalescence occur at the CEP[9]. In some special cases, the CEP taking place at finite temperatures can be tuned to zero temperature by applying a magnetic field, pressure, or adjusting chemical composition[10], yielding a quantum critical point with a diverging susceptibility at zero temperature. An example of metamagnetic quantum criticality in metals has been found in the layered $Sr_3Ru_2O_7$ system in a magnetic field[11-14]. This layered magnetic compound associated with itinerant electrons shows a first-order metamagnetic transition which terminates in a CEP at about 1 K when a magnetic field is applied within the layers. It appears that the CEP is tuned to zero temperature at a magnetic field of ~8 T, applied almost perpendicular to the layers. Despite such detailed research on the quantum criticality of itinerant magnets, critical phenomena in the large class of insulating magnets have scarcely been discussed. In an insulating magnet of $BiMn_2O_5$, the existence of a CEP induced by a magnetic field was conjectured but not observed due to an extremely low critical temperature[15, 16]. Furthermore, any giant fluctuation phenomena corresponding to critical opalescence have not been observed at any CEP's of magnets. To address this issue, we have investigated the metamagnetism of the insulating multiferroic $HoMnO_3$ (h-HMO).

Hexagonal rare-earth manganites, $REMnO_3$ (RE=Sc, Y, Ho, …, Lu), possessing $P6_3cm$ symmetry, have been extensively investigated due to their unique improper ferroelectricity arising from the structural trimerization[17-19] and the co-occurrence of ferroelectricity with antiferromagnetism[20-23]. In addition, recently, fascinating vortex and antivortex networks associated with the interlocked antiphase and ferroelectric domains were also found[24]. The detailed magnetic structures of the Mn spins in h-HMO have been well established, with the planar triangles of spins initially forming a 120° antiferromagnetic configuration below $T_N$=75 K, that flops at $T_{SR}$≈37 K into a different 120° arrangement, and into yet a third type of 120° spin configuration below $T_{Ho}$≈5 K when the $Ho^{3+}$ spins develop spontaneous order[25-28]. In our investigation, at ~2 K where both the $Mn^{3+}$ and $Ho^{3+}$ spins are ordered, we observe an unexpected divergence in the dissipative magnetic susceptibility. This emergence of a singular anomaly is identified as a CEP in *H-T* phase diagram, above which the transition is continuous and evolves smoothly as it crosses over from one phase to the other. To explain this behavior, we consider the Dzyaloshinskii-Moriya (DM) interactions between neighboring $Ho^{3+}$ and $Mn^{3+}$, which enables an understanding of the full magnetic structure consistent with the observed isothermal metamagnetic-jumps in the magnetization and the real and dissipative susceptibilities. We argue that the critical fluctuations associated with the CEP originate between two magnetic phases with the same symmetry. Interestingly, the magnetic-field dependence of thermal resistivity shows similar anomalous behavior as the real part of the magnetic susceptibility, suggesting strong spin-phonon coupling accompanying the magnetic critical fluctuations.

Single crystals of $HoMnO_3$ were grown by the floating zone method. DC magnetization, *M*, was obtained using a SQUID magnetometer (Quantum Design MPMS) and AC magnetic susceptibility was taken in a Quantum Design PPMS with the excitation of 5 Oe at $f$ =1 kHz. In order to complete the *T-H* phase diagram, the temperature dependence of the dielectric constant, $\varepsilon'$, and electric polarization, *P*, were measured using an LCR meter at $f$ =1 kHz and attained by the integration of the pyroelectric current measured in an electrometer with the temperature variation of 4 K/min after poling from high temperature, respectively. The thermal conductivity, $\kappa$, was measured utilizing the conventional steady-state method. The neutron diffraction measurements were carried out using a 7 T superconducting magnet and $^3$He insert on the BT-9 triple-axis spectrometer, with both monochromator and analyzer set for 14.7 meV.

Figures 1 (a) and (b) depict the crystallographic structure of h-HMO seen from the *c* axis and the direction perpendicular to the *c* axis, respectively. The structure comprises triangular lattice layers of $Mn^{3+}O_5$ polyhedra which are tilted to form the trimerization of $Mn^{3+}$ ions in each layer. Ferroelectric polarization emerging along the *c* axis results from the trimer-induced opposite displacements of $Ho^{3+}$ ions with non-equal portions (downward displacements of 2/3 4*b*-site $Ho^{3+}$ ions and upward displacements of 1/3 2*a*-site $Ho^{3+}$ ions)[17], occurring at $T_c \approx$ 1370 K[29]. The five different phases are apparent in the series of metamagnetic transitions of isothermal magnetization with both ramping up and down curves under the magnetic field along the *c* axis measured at 2.2 K up to 3.0 T (Fig. 1(c)) [28, 30-32]. We label the magnetic phases as A1, $A1_{0°}$, $A1_{30°}$, $A1_{60°}$, and A2 in order of increasing magnetic field. Distinct magnetic hysteresis was observed at the transition between the $A1_{0°}$ and $A1_{30°}$ phases. Note that the phases of $A1_{30°}$ and $A1_{60°}$ are denoted in red and blue, respectively, because the transition from $A1_{30°}$ to $A1_{60°}$ at which magnetic symmetry group is conserved[33] is special, and our discussion now will mainly focus on this transition.

In order to examine the metamagnetic transitions in $M(\mu_0 H)$, the imaginary and real parts of the AC magnetic susceptibility under magnetic field along the *c* axis, $\chi''(\mu_0 H)$ and $\chi'(\mu_0 H)$, were measured in the low $T$ regime up to 4 T (Figs. 2(c) and (d)), and *H*-*T* contour plots between 2 K and 6.5 K were constructed (Figs. 2(a) and (b)). The plots reveal that a dramatic anomaly, indicative of a strong divergence of $\chi''(H)$, emerges at $\mu_0 H \approx 2.0$ T and $T \approx 2.2$ K, while a broader anomalous feature at $\mu_0 H \approx 2.0$ T and $T \approx 2.3$ K appears in the plot of $\chi'(\mu_0 H)$. This singularity turns out to be a magnetic CEP. The highest values of $\chi''(\mu_0 H)$ and $\chi'(\mu_0 H)$ are colored black to depict the CEP. The phase boundaries in the intensity plot of $\chi'(\mu_0 H)$ are represented as grey dotted curves relevant to the distinct magnetic phases (Fig. 2(b)), indicated in the metamagnetic transitions in $M(\mu_0 H)$ (Fig. 1(c)). For the boundary between the $A1_{30°}$ and $A1_{60°}$ phases, the white solid curve drawn below the black singular anomaly indicates the first order transition line that ends in the CEP and the white dotted curve above the point denotes the crossover regime. Figures 2 (c) and (d) display the 3D plots of $\chi''(\mu_0 H)$ and $\chi'(\mu_0 H)$ at various temperatures at $T$= 2.0 K - 4.5 K. The data reveal four different magnetic transitions, but the drastic change of the peaks occurs near the CEP.

Despite the considerable work to determine the complete spin structures for all the distinct magnetic phases below $T_{Ho}$ under the magnetic field along the *c*-axis, the exact configurations have not yet been fully clarified because of the complexity of the system and experimental difficulty in separating magnetic contributions of both $Ho^{3+}$ sites and $Mn^{3+}$ magnetic moments. Fully established magnetic structures are only given for the $Mn^{3+}$ spin ordering in the A1 phase at zero-field, and in the A2 phase at high field[25-28, 30, 33, 34]. $Mn^{3+}$ spin ordering in the A2 phase can be achieved by clockwise rotation of each $Mn^{3+}$ spin in the A1 phase by 90°[33]. Thus, it naturally leads to the plausible assumption that $Mn^{3+}$ spins in the intermediate phases lie somewhere between 0° ($Mn^{3+}$ spins in A1 phase) and 90° ($Mn^{3+}$ spins in A2 phase). Taking into account the crystal structure of h-HMO, the most plausible configurations correspond to 30° and 60° rotations relative to 0°. To accomplish the magnetic structures, first we have constructed conceivable schemes of $Mn^{3+}$ spin configurations of the three intermediate phases, i.e., $A1_{0°}$ (where $Mn^{3+}$ spins order in the same way as A1 but $Ho^{3+}$ spins in 2*a* sites are magnetically polarized), $A1_{30°}$ and $A1_{60°}$ (see the SI for details). Then, we have established the $Ho^{3+}$ moments of the five different phases, oriented by the antisymmetric exchange fields from neighboring $Mn^{3+}$ ions[35-39]. The directions and the strength of effective magnetic fields originating from the Dzyaloshinskii-Moriya (DM) interactions are discussed in the SI in detail. Note that the exact strength of the effective fields and the magnitudes of the induced $Ho^{3+}$ moments could not be easily quantified. However, our analysis manifests that upon increasing magnetic fields, the induced $Ho^{3+}$ moments increase gradually, consistent qualitatively with the identified magnetic phases in the $M$($H$) curves shown in Fig. 1(c).

Figure 3(a) displays the planar magnetic structure of h-HMO within a magnetic unit cell for the $A1_{30°}$ and $A1_{60°}$ phases. The arrows with and without outlines correspond to $z$=1/2 and $z$=0 $Mn^{3+}$ layers, respectively, and the red and blue arrows indicate a relative orientation of 30° to each other for the $Mn^{3+}$ spins in the $A1_{30°}$ and $A1_{60°}$ phases, respectively. It is apparent that the transition from the $A1_{30°}$ to $A1_{60°}$ phase does not involve a symmetry change[33], leading to a CEP which terminates the line of first order transitions in the phase diagram. For comparison, we note that the angle for the canted moments of $Cu^{2+}$ spins in $La_2CuO_4$ that give rise to the weak ferromagnetism is about 10 % of the angle of rotational distortion for oxygen octahedra[40]. Similarly, the rotational angle of the $Mn^{3+}O_5$ polyhedra associated with the trimerization in h-HMO is calculated as ~11°, and thus it is expected that the tilting angle of $Mn^{3+}$ spins along the *c*-axis in the A2 phase is small (~1°). As a consequence, the in-plane angular variation of the spins by 30° between the $A1_{30°}$ to $A1_{60°}$ phases is accompanied by a small out-of-plane variation of $Mn^{3+}$ spins of probably only ~0.4°. We refer to the spin fluctuations at the CEP as 'small-helix-angle screw type', as illustrated in Fig. 3(b). The in-plane component of the spin fluctuation dominates, since the out-of-plane angle is estimated to be less than half a degree. Therefore the in-plane component is likely responsible for observed giant dissipation in the AC magnetic susceptibility. Figure 3(c) shows the comparison of normalized $\chi''(\mu_0 H)$ and *H*-derivative of magnetization, $dM/d\mu_0 H$, both measured at 2.2 K. In spite of the presence of only a small peak of $dM/d\mu_0 H$, a striking divergence of dissipation at 2.05 T, which exceeds the value at 3.0 T by more than 200 times, develops at the transition between $A1_{30°}$ and $A1_{60°}$ phases.

The first order nature of the transition below the temperature of the CEP is well evidenced in the temperature dependent plot of the full-width-half-maximum (FWHM) values estimated from the $dM/d\mu_0 H$ peaks, and the shift of $\chi'(\mu_0 H)$ between ramping up and down curves, $\Delta\mu_0 H(T)$ (Fig. 4(a)). The abrupt decrease of $\Delta\mu_0 H(T)$ and the sudden increase of FWHM at ~2.3 K clearly delineates the boundary between the first order transition line and the smooth crossover on either side of the CEP. Another important discovery was the substantial suppression of thermal conductivity[41] across the CEP. It turns out that the plot of the reciprocal of the magnetic-field dependent (*H*//*c*) thermal conductivity, $\kappa$, measured at 2.0 K, behaves similarly to $\chi'(\mu_0 H)$ (Fig. 4(b)). This suggests that the strong field dependence of the thermal conductivity accompanying the magnetic critical fluctuations results from strong spin-phonon coupling. The huge susceptibility dissipation suggests that the anomaly of thermal conductivity also should be significant, however, the thermodynamic non-equilibrium state formed during the measurement allows only a small fraction of the sample to exhibit critical fluctuations.

In the immediate vicinity of a CEP, critical fluctuations may be governed by the usual scaling theory, with a divergent susceptibility at the CEP[42, 43]. However, the critical behavior near the CEP in h-HMO appears to be unique, suggesting that this system may belong to a new universality class. We extract the critical exponents from the magnetization and the real part of the AC susceptibility with varying magnetic fields and temperature near the CEP, respectively (Figs. 4(c) and (d)). A scaling factor, δ, can be evaluated from the plot of $M(\mu_0 H)$-$M(\mu_0 H_{crit.}$=2.05±0.025 T) that is proportional to $(\mu_0 H-\mu_0 H_{crit.})^{1/\delta}$ near the CEP, and γ from the plot of $\chi'$ in $\mu_0 H$=2.05 T with respect to $(T-T_{crit.})^{-\gamma}$ where $T_{crit.}$ is 2.3±0.05 K. The observed scaling exponents of (δ, γ)=(1.28±0.23, 0.58±0.03) deviate from those for the 3D XY-model, anticipated for h-HMO, where (δ, γ)≈(4.8, 1.3) [44, 45]. These unusual values of δ and γ may reflect the unique collective nature of the giant small-helix-angle screw-type spin fluctuations (with the configuration of in-plane 120° antiferromagnetic and tiny *c*-direction ferromagnetic moments) between the $A1_{30°}$ and $A1_{60°}$ magnetic phases with the identical magnetic symmetry.

In summary, we report the discovery of a divergent dissipative magnetic susceptibility at $\mu_0 H$≈2.0 T and $T$≈2.2 K in an insulating metamagnet of hexagonal $HoMnO_3$. The identification of several low-temperature metamagnetic phases regarding both $Ho^{3+}$ and $Mn^{3+}$ spins reveals that the singular anomaly in the $H$-$T$ phase diagram turns out to be a critical endpoint. The giant critical fluctuations at the endpoint, mimicking critical opalescence in fluids, stem from the zero energy motion under AC magnetic excitation for the two magnetic phases with the same symmetry, and can be described as a small-helix-angle screw type. In addition, we observed a significant suppression of thermal conductivity at the critical endpoint, indicating strong spin-phonon coupling accompanying the magnetic critical fluctuations. These intriguing results reveal novel physics in the insulating hexagonal manganites and should invigorate new explorations of critical phenomena in magnetic insulators.

## Figure Captions

**FIG. 1** (color online). (a) and (b) Views of the crystal structure of hexagonal $HoMnO_3$ (*$P6_3cm$*) from the *c* axis and perpendicular to the *c* axis. Green, purple, blue and yellow spheres represent $Ho^{3+}$(at 2*a* sites), $Ho^{3+}$(at 4*b* sites), $Mn^{3+}$, and $O^{2-}$ ions, respectively. The dark yellow box with the cross-section parallelogram along the *c* axis designates the crystallographic unit cell. (c) Isothermal magnetization with both ramping up and down measurements along the *c*-axis at 2.2 K. The five distinct magnetic phases are labeled in order of increasing magnetic fields ($A1$-$A1_{0°}$-$A1_{30°}$-$A1_{60°}$-$A2$).

**FIG. 2** (color online). (a) and (b) *H-T* contour plots in the low temperature regime up to 4 T (*H//c*) obtained from the imaginary and real parts of the AC magnetic susceptibility, $\chi''(\mu_0 H)$ and $\chi'(\mu_0 H)$, respectively. The phase boundaries in the intensity plot of $\chi'(\mu_0 H)$ are depicted as grey dotted curves, and the white solid-curve and white dotted-curves that divide the $A1_{30°}$ and $A1_{60°}$ phases indicate the first order transition line below the CEP and the crossover above the CEP, respectively. (c) and (d) Magnetic-field dependence (*H//c*) of the imaginary and real parts of the AC magnetic susceptibility at various temperatures between 2.0 K and 4.5 K. The solid black points in all four figures locate the CEP. Note that 1 emu/(mol Oe) = $4\pi\times10^{-6}$ $m^3$/mol.

**FIG. 3** (color online). (a) Planar magnetic structure of h-HMO within a magnetic unit cell for the $A1_{30°}$ and $A1_{60°}$ phases. Red (blue) arrows correspond to $Mn^{3+}$ spins in the $A1_{30°}$ ($A1_{60°}$) phase. The arrows with (without) outlines indicate $z$=1/2 ($z$=0) $Mn^{3+}$ layers. Note that the transition between the $A1_{30°}$ and $A1_{60°}$ phase does not involve any symmetry change. (b) Schematic presenting large in-plane spin fluctuations of the small-helix-angle screw type. Light grey, red, blue, light purple arrows indicate $Mn^{3+}$ spins of A1, $A1_{30°}$, $A1_{60°}$, and A2 phases, respectively. Helix angle, $\theta_{helix}\approx0.4$ °, is exaggerated in comparison with the in-plane variation of $\theta_{in\text{-}plane}\approx30$ ° for clarity. (c) $\chi''(\mu_0 H)$ normalized to the value at 3.0 T (2.2 K) and $H$-derivative of the magnetization (2.2 K). At the CEP, $\chi''$(2.05 T, 2.2 K)/$\chi''$(3 T, 2.2 K)≈235.

**FIG. 4** (color online). (a) Temperature dependence of the full-width-half-maximum (FWHM) (solid circles) estimated from the d$M$/d$\mu_0 H$ peaks, and the shift of $\chi'(\mu_0 H)$ between ramping up and down curves (open circles), both occurring at the boundary between $A1_{30°}$ and $A1_{60°}$ phases. (b) Magnetic field dependence ($H//c$) of the inverse thermal conductivity (2.0 K) and $\chi'(\mu_0 H)$ normalized to the value at 3.0 T (2.3 K) (c) and (d) Critical exponents of (δ, γ)≈(1.28±0.23, 0.58±0.03) obtained from (c) $M(\mu_0 H)$-$M(\mu_0 H_{crit.})$ vs. $(\mu_0 H$-$\mu_0 H_{crit.})/(\mu_0 H_{crit.})$ where $\mu_0 H_{crit.}$=2.05±0.025 T, and (d) ($T$-$T_{crit.}$)/($T_{crit.}$) dependence of the real part of the AC magnetic susceptibility, $\chi'$, in $\mu_0 H$=2.05 T with $T_{crit.}$=2.3±0.05 K. Uncertainties represent one standard deviation.

## References

[1] E. Dagotto, Science **309**, 257 (2005).

[2] J. Zaanen, Nature **448**, 1000 (2007).

[3] H. E. Stanley, *Introduction to Phase Transitions and Critical Phenomena* (Clarendon Press, Oxford, 1971).

[4] C. Domb, *The Critical Point: A historical introduction to the modern theory of critical phenomena* (Taylor & Francis, 1996).

[5] A. Barducci *et al.*, Phys. Lett. B **231**, 463 (1989).

[6] R. V. Gavai, and S. Gupta, Phys. Rev. D **71**, 114014 (2005).

[7] R. A. Lacey *et al.*, Phys. Rev. Lett. **98**, 092301 (2007).

[8] O. Rixe, and T. Fojo, Clin. Cancer Res. **13**, 7280 (2007).

[9] E. Gopal, Resonance **5**, 37 (2000).

[10] G. Aeppli, and Y. A. Soh, Science **294**, 315 (2001).

[11] S. A. Grigera *et al.*, Science **294**, 329 (2001).

[12] S. A. Grigera *et al.*, Science **306**, 1154 (2004).

[13] R. A. Borzi *et al.*, Science **315**, 214 (2007).

[14] A. W. Rost *et al.*, Science **325**, 1360 (2009).

[15] J. W. Kim *et al.*, Proc. Natl. Acad. Sci. **106**, 15573 (2009).

[16] G. S. Jeon *et al.*, Phys. Rev. B **79**, 104437 (2009).

[17] B. B. Van Aken *et al.*, Nat. Mater. **3**, 164 (2004).

[18] C. J. Fennie, and K. M. Rabe, Phys. Rev. B **72**, 100103 (2005).

[19] S.-W. Cheong, and M. Mostovoy, Nat. Mater. **6**, 13 (2007).

[20] M. Fiebig *et al.*, Nature **419**, 818 (2002).

[21] B. Lorenz *et al.*, Phys. Rev. Lett. **92**, 087204 (2004).

[22] B. Lorenz *et al.*, Phys. Rev. B **70**, 212412 (2004).

[23] P. G. Radaelli, and L. C. Chapon, Phys. Rev. B **76**, 054428 (2007).

[24] T. Choi *et al.*, Nat. Mater. **9**, 253 (2010).

[25] O. P. Vajk *et al.*, Phys. Rev. Lett. **94**, 087601 (2005).

[26] P. J. Brown, and T. Chatterji, J. Phys.: Condens. Matter **18**, 10 085 (2006).

[27] S. Nandi *et al.*, Phys. Rev. Lett. **100**, 217201 (2008).

[28] P. Brown, and T. Chatterji, Phys. Rev. B **77**, 104407 (2008).

[29] S. C. Chae *et al.*, Phys. Rev. Lett. **108**, 167603 (2012).

[30] O. P. Vajk *et al.*, J. Appl. Phys. **99**, 08E301 (2006).

[31] B. Lorenz *et al.*, Phys. Rev. B **71**, 014438 (2005).

[32] F. Yen *et al.*, Phys. Rev. B **71**, 180407 (2005).

[33] M. Fiebig, T. Lottermoser, and R. V. Pisarev, J. Appl. Phys. **93**, 8194 (2003).

[34] I. Munawar, and S. H. Curnoe, J. Phys.: Condens. Matter **18**, 9575 (2006).

[35] T. Moriya, Phys. Rev. **120**, 91 (1960).

[36] X. Fabrèges *et al.*, Phys. Rev. B **78** (2008).

[37] K. Matan *et al.*, Phys. Rev. B **83** (2011).

[38] K. Matan *et al.*, Phys. Rev. Lett. **96** (2006).

[39] Y. Geng *et al.*, Nano Letters **12**, 6055 (2012).

[40] S. W. Cheong, J. D. Thompson, and Z. Fisk, Phys. Rev. B **39**, 4395 (1989).

[41] X. M. Wang *et al.*, Phys. Rev. B **82**, 094405 (2010).

[42] A. Millis *et al.*, Phys. Rev. Lett. **88**, 217204 (2002).

[43] U. Adem *et al.*, J. Phys.: Condens. Matter **21**, 496002 (2009).

[44] L. P. Kadanoff *et al.*, Reviews of Modern Physics **39**, 395 (1967).

[45] F. Kagawa, K. Miyagawa, and K. Kanoda, Nature **436**, 534 (2005).

**Acknowledgements**

Work at Rutgers was supported by DOE DE-FG02-07ER46328. Y.J.C. was supported partially by the Yonsei University Research Fund of 2011. N.L. was supported partially by the Yonsei University Research Fund of 2012.

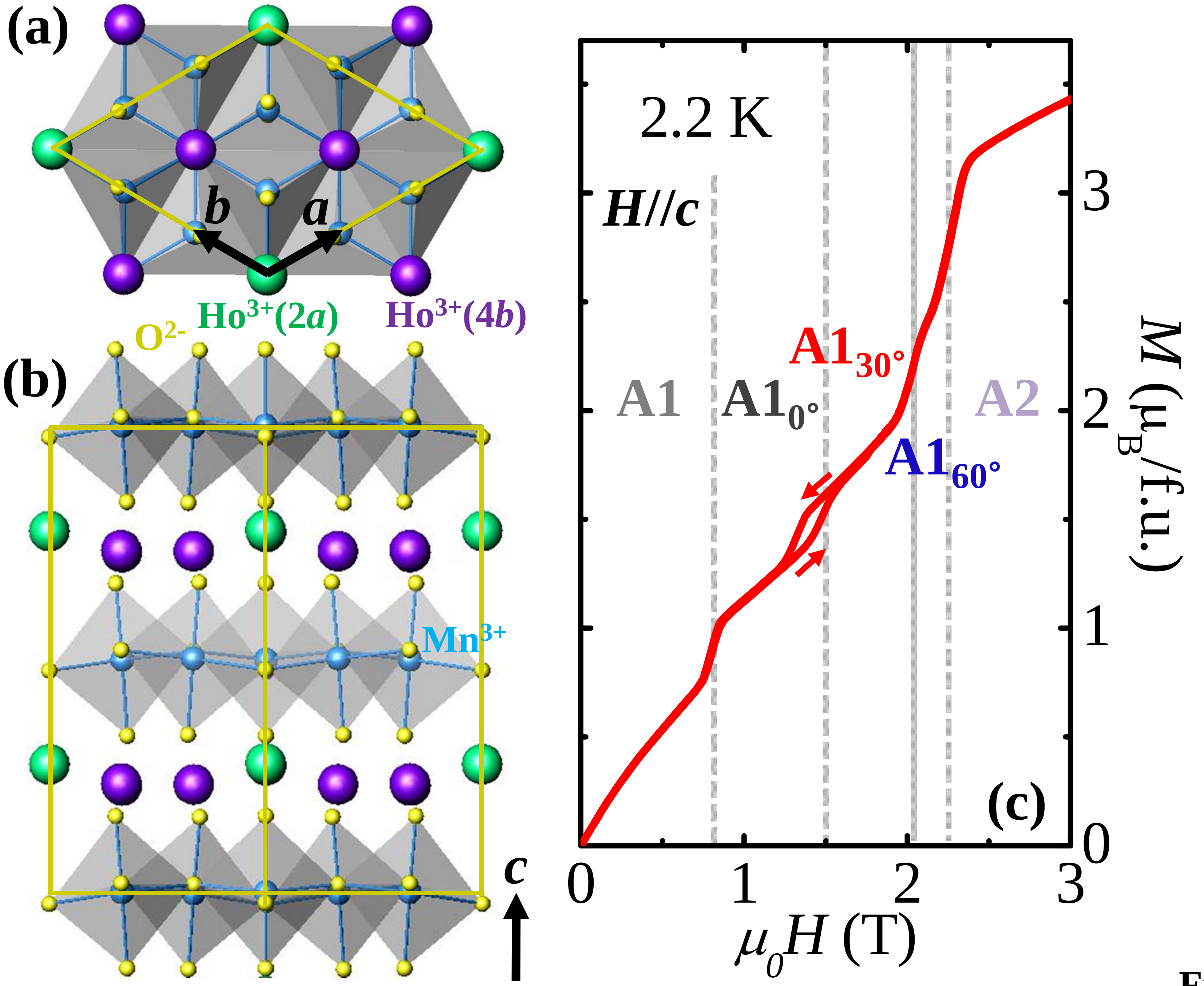

Fig. 1

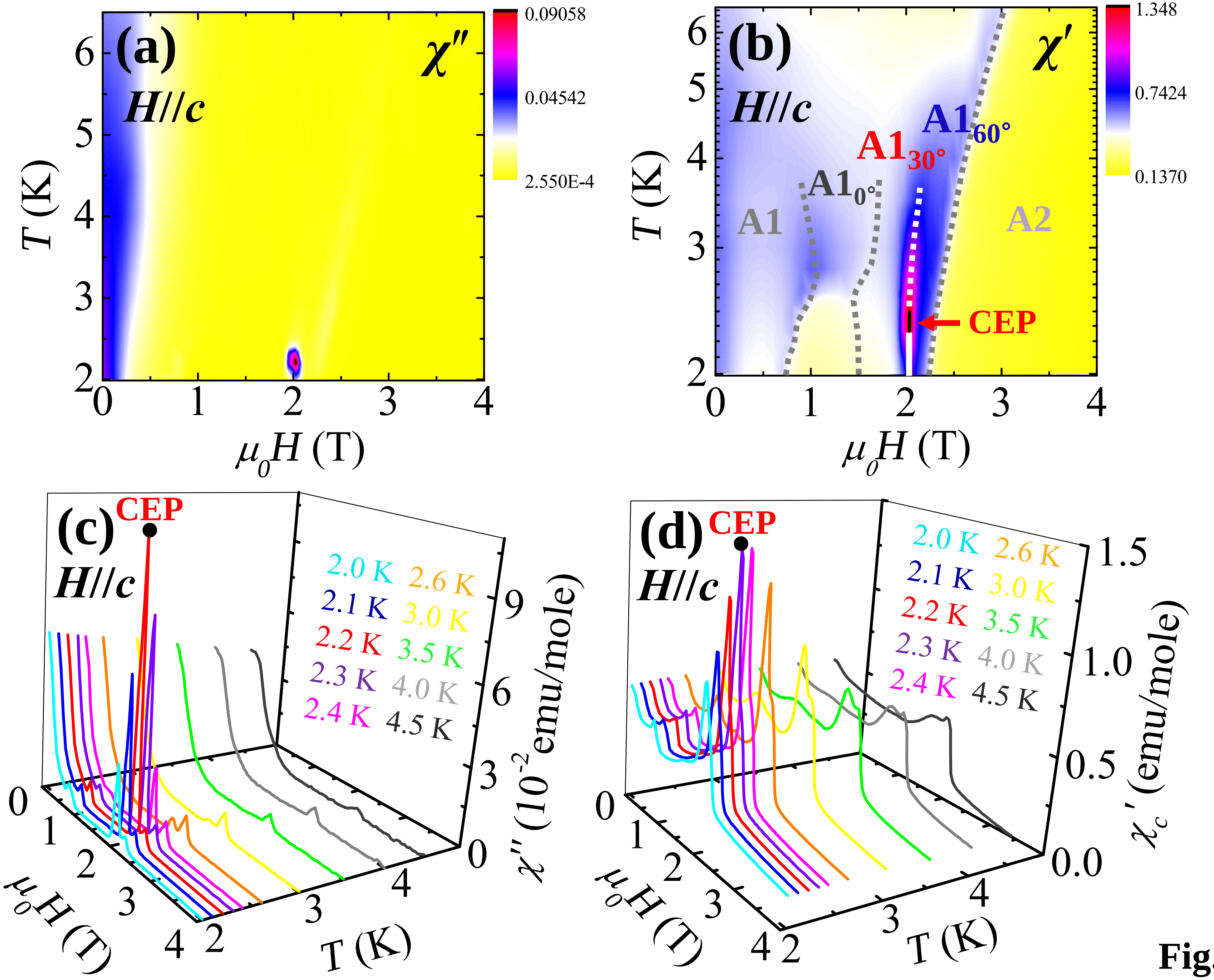

**Fig. 2**

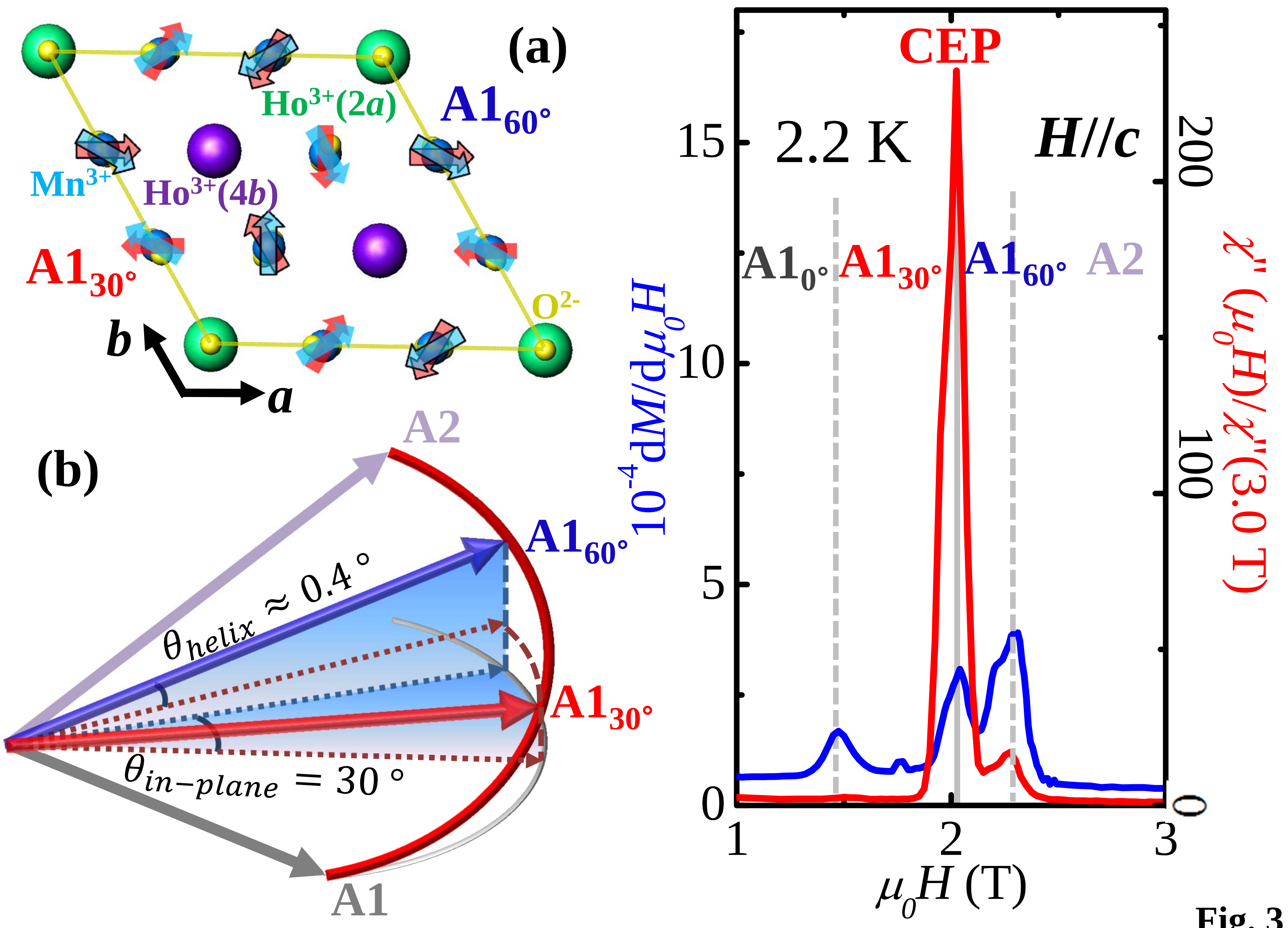

Fig. 3

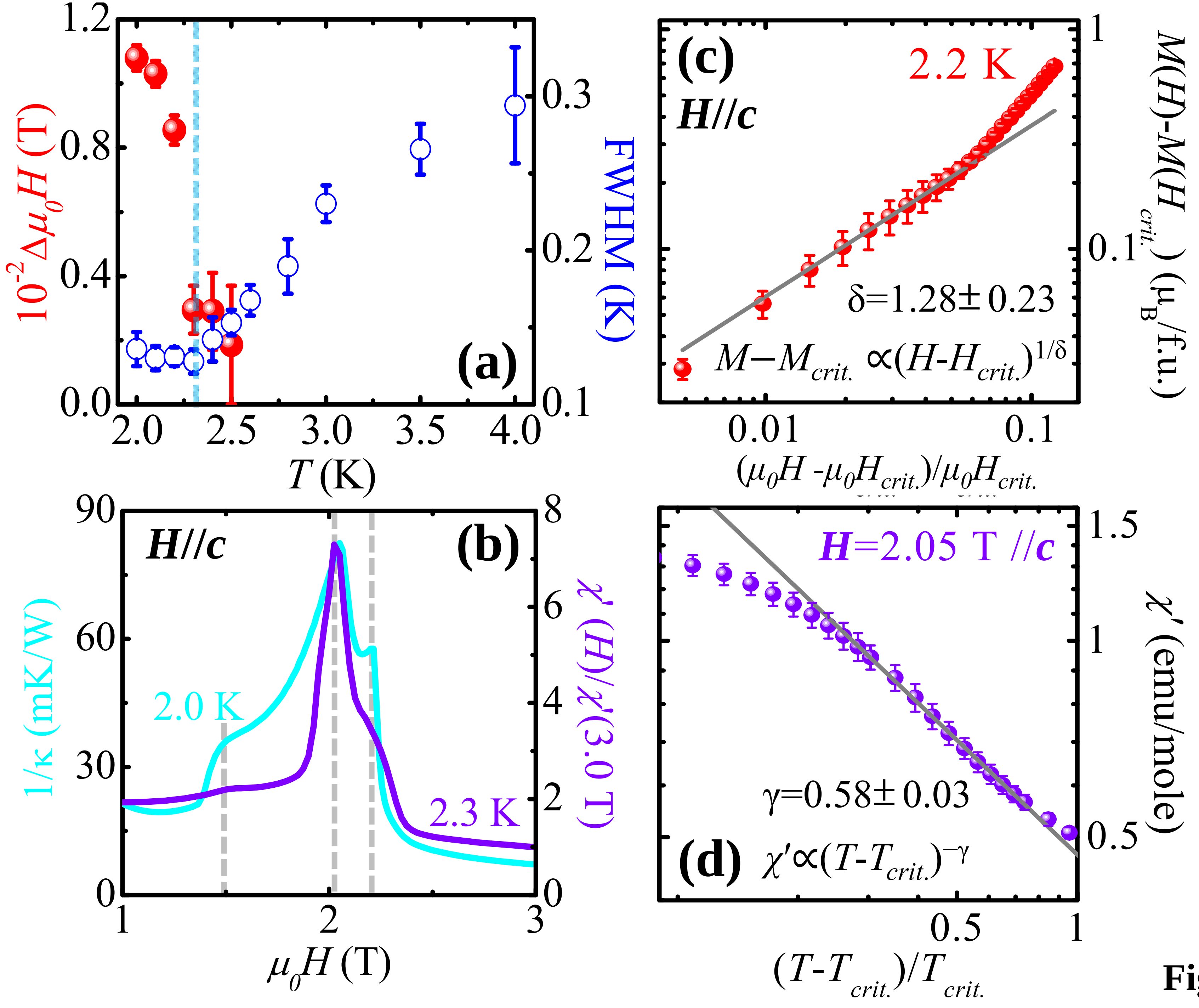

**Fig. 4**

Supplementary Information for

# Giant Magnetic Fluctuations at the Critical Endpoint in Insulating $HoMnO_3$

Y. J. Choi[1,2], N. Lee[1,2], P. A. Sharma[1], S. B. Kim[1], O.P. Vajk[3], J. W. Lynn[3], Y. S. Oh[1], and S-W. Cheong[1]

[1] *Rutgers Center for Emergent Materials and Department of Physics & Astronomy, Rutgers University, Piscataway, New Jersey, 08854, USA*

[2] *Department of Physics and IPAP, Yonsei University, Seoul 120-749, Korea*

[3] *NIST Center for Neutron Research, National Institute of Standards and Technology, Gaithersburg, Maryland 20899, USA*

**DM (Dzyaloshinskii-Moriya) interactions and induced $Ho^{3+}$ magnetic moments**

Figure S1 displays the 3D crystallographic structure (*P6₃cm*) and selective DM exchange paths of hexagonal $HoMnO_3$ (h-HMO). The DM exchange paths between $Ho^{3+}$ and $Mn^{3+}$ ions via apical (planar) $O^{2-}$ ions, are represented by dotted blue (orange) arrows. The shaded blue (orange) regions are the planes where $Mn^{3+}$-$O^{2-}$-$Ho^{3+}$ paths lie. The DM vector in each path can be determined by the rotational axis of the $Mn^{3+}$-$O^{2-}$ bond distortion applying the right-hand rule. The direction of a DM vector is shown in the white symbol which indicates one of the perpendicular directions to the shaded planes. The DM interactions that describe $Mn^{3+}$-$O^{2-}$-$Ho^{3+}$ exchange paths are given by $H_{DM} = -\sum \vec{D} \cdot (\vec{S}_{Mn} \times \vec{S}_{Ho})$, where $\vec{D}$, $\vec{S}_{Mn}$ and $\vec{S}_{Ho}$ are DM vector, $Mn^{3+}$ spin, and $Ho^{3+}$ spin, respectively[1]. The summation includes six nearest neighbors of $Mn^{3+}$ spins (three from the $Mn^{3+}$ trimer in $z$=0 layer and the other three in $z$=1/2 layer) with respect to each $Ho^{3+}$ ion at 2$a$ or 4 $b$ site. Accompanied by the detail that the 2$a$ and 4$b$ sites of $Ho^{3+}$ ions are not in the same plane, bold (regular) letter for DM vector, $D$, indicates stronger (weaker) DM interaction resulting from the shorter (longer) path. The permutation relation leads the interactions to $H_{DM} = -\sum \vec{S}_{Ho} \cdot (\vec{D} \times \vec{S}_{Mn}) = -\vec{S}_{Ho} \cdot \vec{H}_{eff}$, where $\vec{H}_{eff} = \sum \vec{D} \times \vec{S}_{Mn}$ is the effective magnetic fields which polarize the $Ho^{3+}$ spins[2]. The step-like features associated with the metamagnetic transitions in the $M(H)$ curve at low $T$ (Fig. 1(c) in the main text) are classified into five different magnetic phases regarding $Mn^{3+}$ and induced $Ho^{3+}$ magnetic moments, sequentially ordered from low to high magnetic-field phases (A1-$A1_{0°}$-$A1_{30°}$-$A1_{60°}$-A2 phase).

The detailed planar magnetic structures of h-HMO for the five different phases under magnetic fields are presented in figure S2. The arrows without (with) outlines denote $Mn^{3+}$ spins in the $z$=0 ($z$=1/2) layer, and a dark yellow parallelogram designates the magnetic unit cell. The values of color rings encircling the $Ho^{3+}$ ions represent the directions and the strength of effective magnetic fields obtained by scrutinizing the DM interactions. The details are summarized in table S1. The exact strength of the effective fields and the magnitudes of induced $Ho^{3+}$ moments could not be easily quantified. However, the estimated magnitudes of DM vectors manifest that upon increasing magnetic fields, the induced $Ho^{3+}$ moments increase gradually, consistent qualitatively with the identified magnetic phases in the $M(\mu_0 H)$ curves. Note that S. Nandi *et al.* summarized the previous results where the $Ho^{3+}$ spins presumably belong to the $\Gamma_1$ or $\Gamma_3$ magnetic representation[3]. Our analysis based on the DM interactions coincides with the $\Gamma_1$ representation. Two additional small figures on the right of each planar structure correspond to $z$=0 and $z$=1/2 $Mn^{3+}$ trimer layers around the unit cell and provide the DM vector for one of the three $Mn^{3+}$ spins in each trimer since the three spins lead to the equivalent contribution to the effective magnetic field on the $Ho^{3+}$ moment. The 4b sites of $Ho^{3+}$ moments can be distinguished as two kinds for which the effective fields are separately estimated. In the small figures, the color rings are only shown for the $Ho^{3+}$ ions associated with the dominant contributions of the DM interactions in either of the layers.

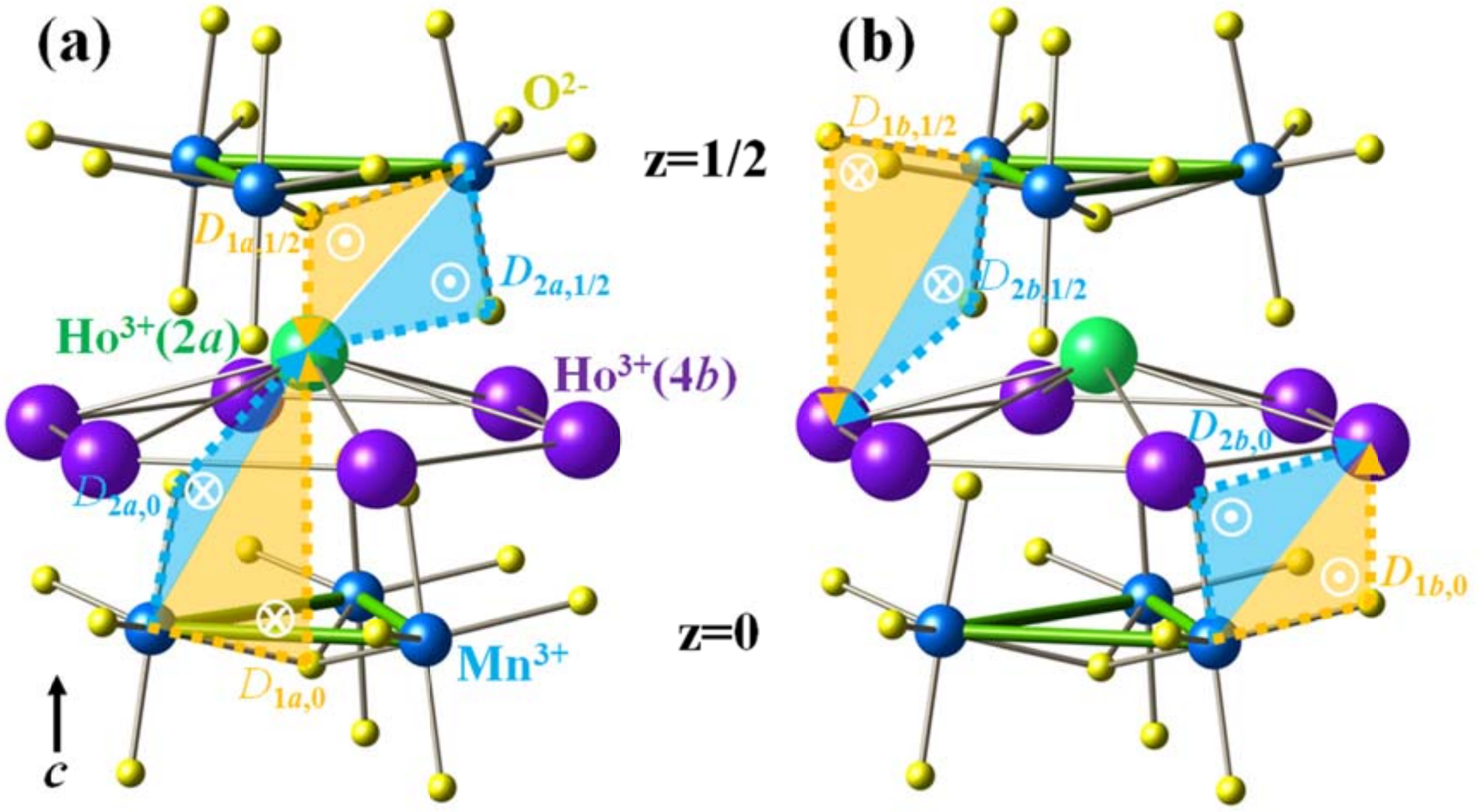

**Figure S1.** (a) and (b) 3D structure cartoons (*$P6_3cm$*) and selective DM exchange paths of hexagonal $HoMnO_3$. Green, purple, blue and yellow spheres represent $Ho^{3+}$(at 2*a* sites), $Ho^{3+}$(at 4b sites), $Mn^{3+}$, and $O^{2-}$ ions, respectively. The DM exchange paths between $Ho^{3+}$ ions at 2*a* (a) and 4*b* (b) sites and $Mn^{3+}$ ions via apical (planar) $O^{2-}$ ions, are represented by dotted blue (orange) arrows. A shaded blue (orange) region is the plane where $Mn^{3+}$-$O^{2-}$-$Ho^{3+}$ path lies.

**(a)** A1

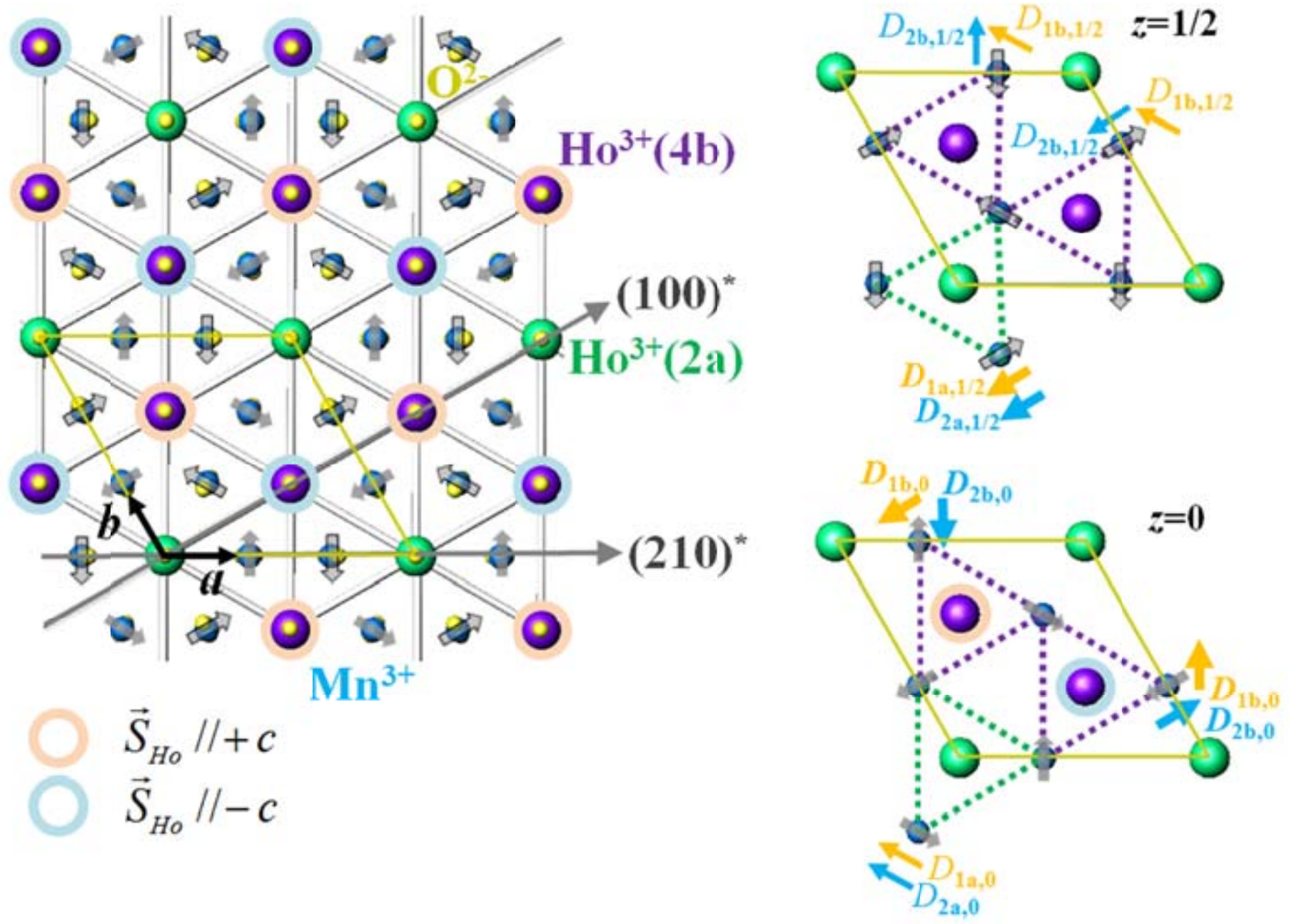

**(b)** $A1_{0^\circ}$

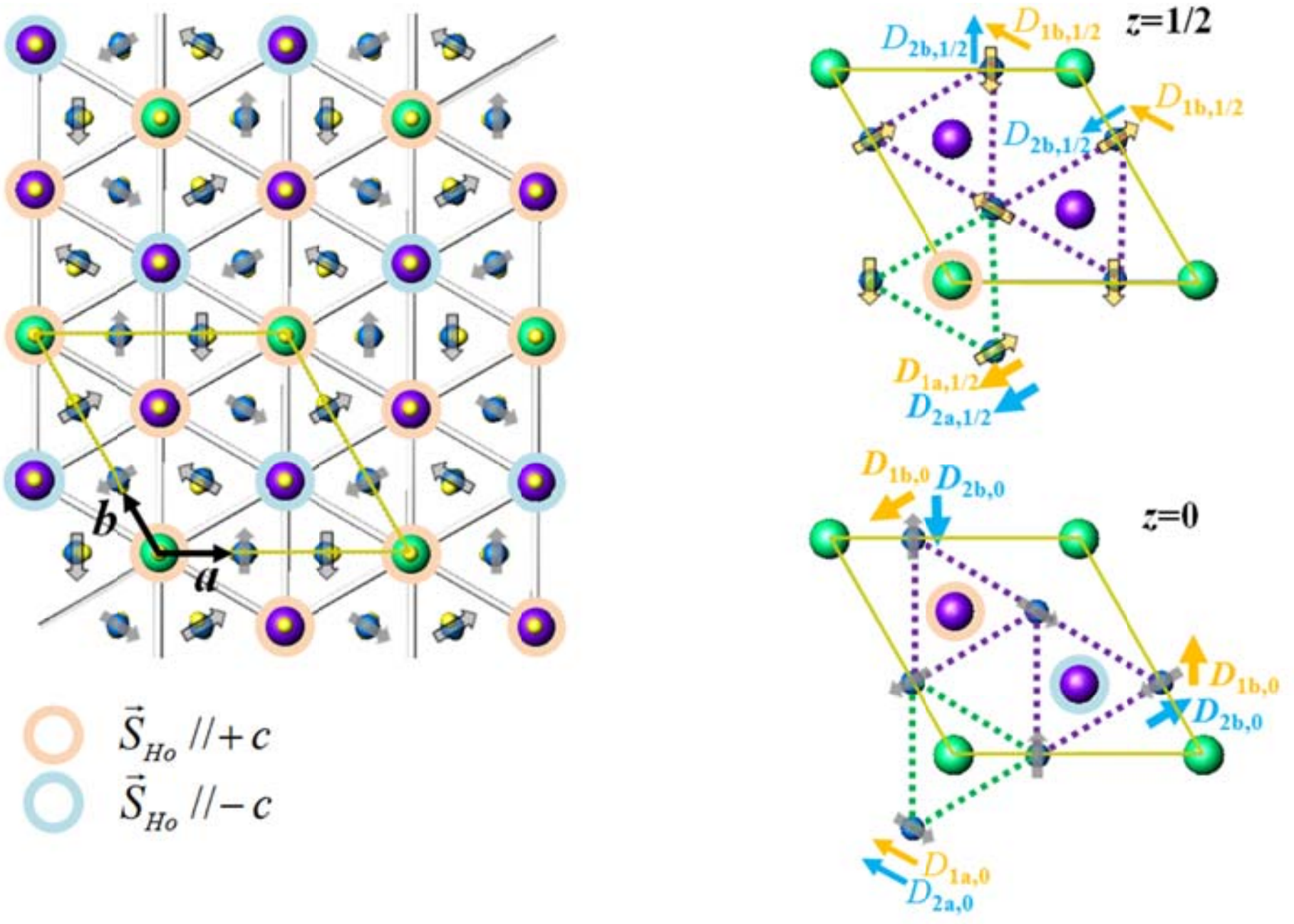

**(c) A1$_{30°}$**

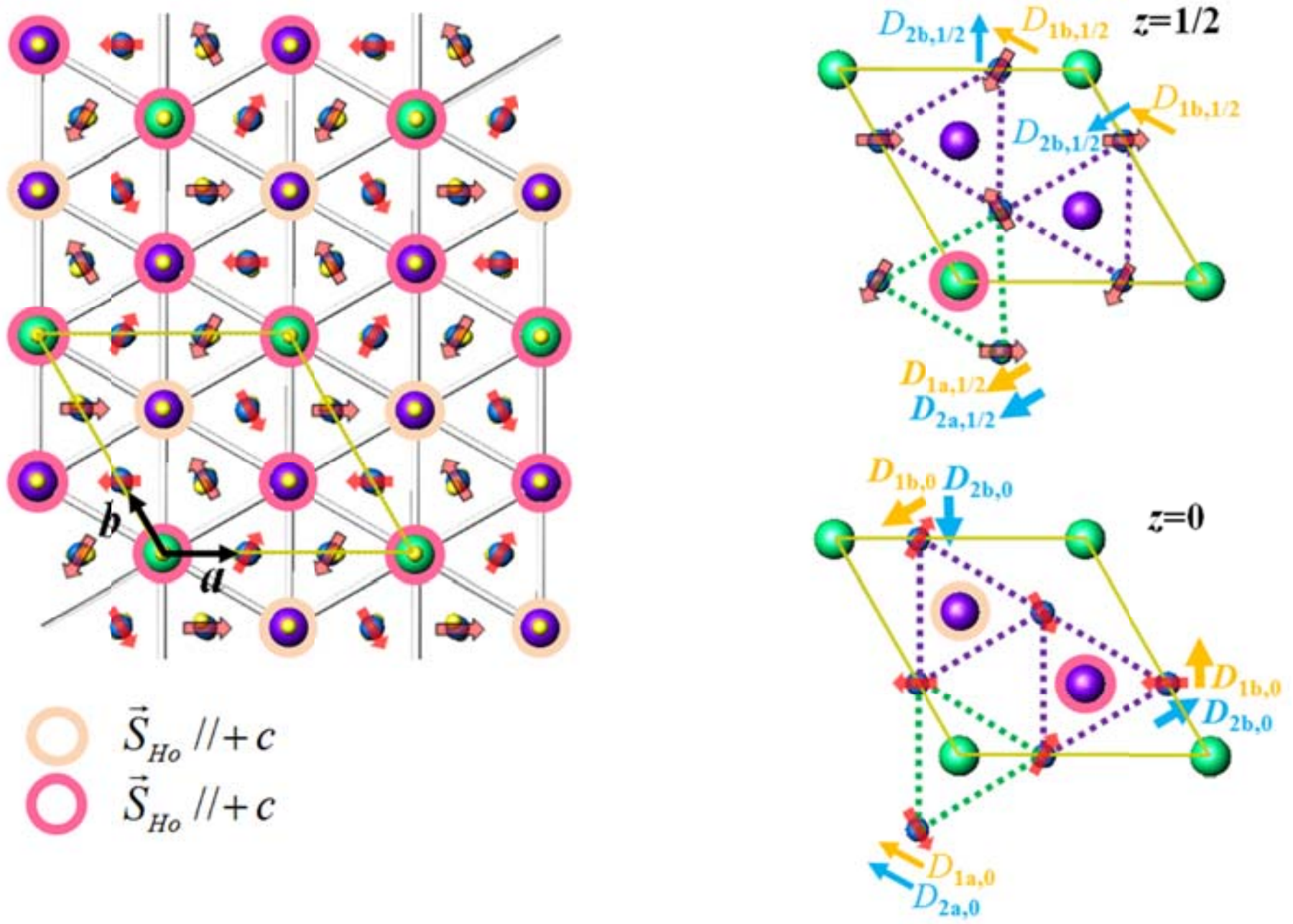

**(d) A1$_{60°}$**

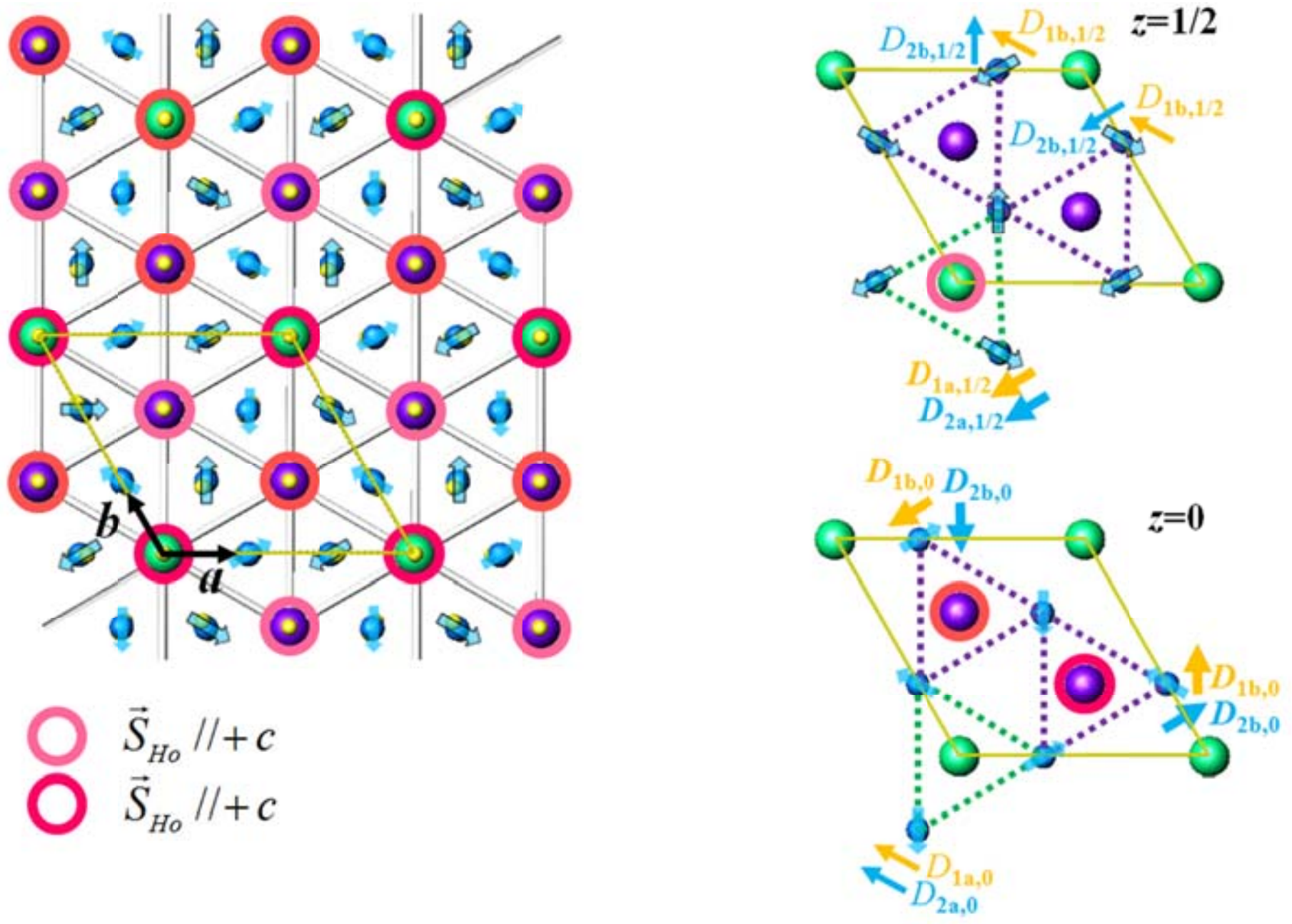

## (e) A2

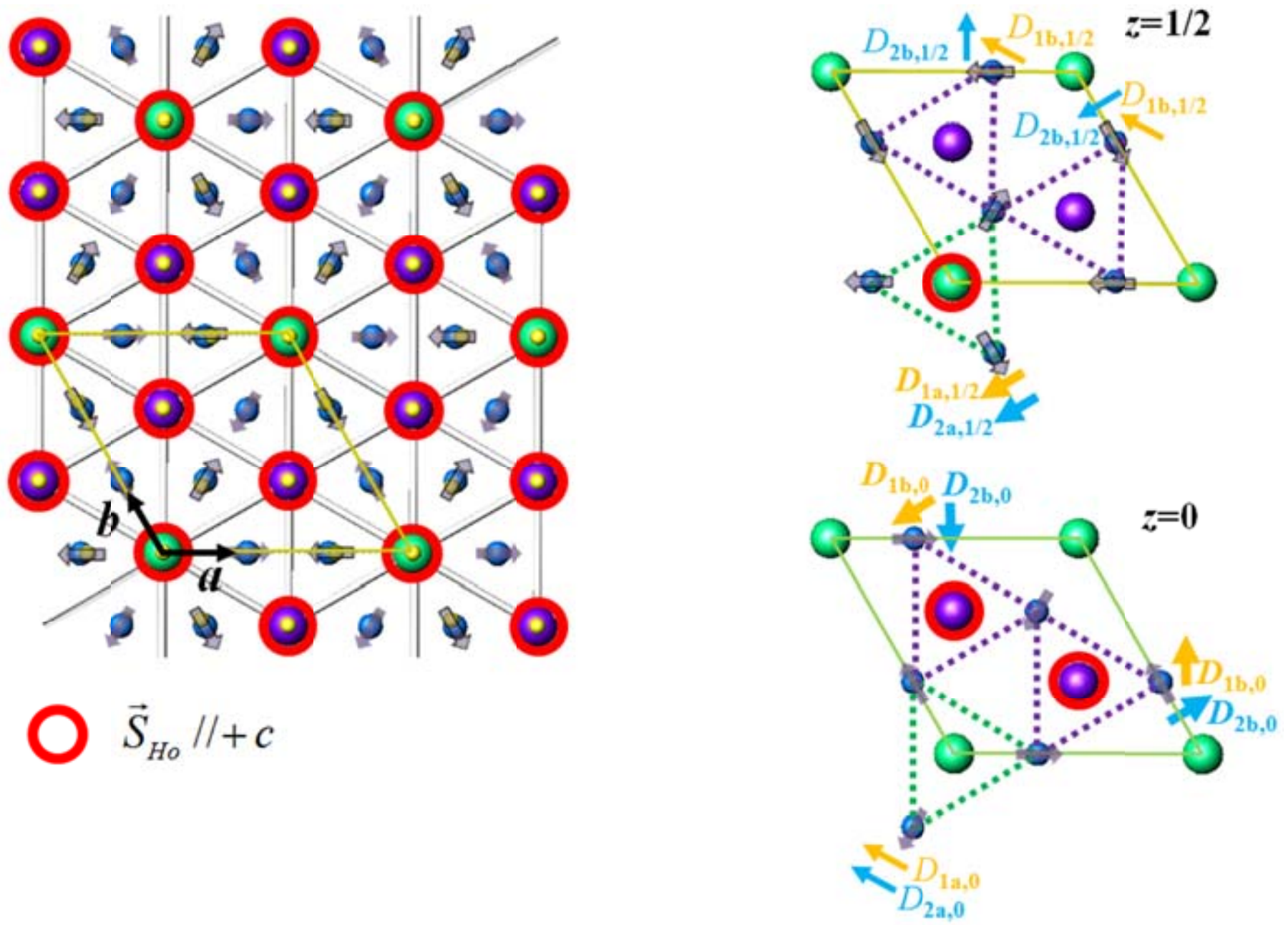

**Figure S2.** (a)-(e) Planar magnetic structures of h-HMO for the five different low-$T$ phases, ordered as A1-A1$_{0^\circ}$-A1$_{30^\circ}$-A1$_{60^\circ}$-A2 from low to high $H$ along the $c$-axis. The arrows without (with) outlines denote $Mn^{3+}$ spins in the $z$=0 ($z$=1/2) layer. The color rings encircling the $Ho^{3+}$ ions represent the directions and the strength of effective magnetic fields. The value of color presents the relative strength of the effective magnetic fields. Two additional small figures on the right of each planar structure correspond to $z$=0 and $z$=1/2 $Mn^{3+}$ trimer layers around the unit cell and show the DM vector for one of the 3 $Mn^{3+}$ spins in each trimer.

**(a)**

| Phase | Sites | $H_{\mathrm{eff}}$ | $Ho^{3+}$ moment | |
|---|---|---|---|---|
| A1 | $Ho_{2a}$ | 0 | 0 | Fig. S2 (a) |
| | $Ho_{4b,1}$ | $\frac{3\sqrt{3}}{2}(\boldsymbol{D}_{1b,0}+D_{1b,1/2})\hat{z}$ | + | |
| | $Ho_{4b,2}$ | $-\frac{3\sqrt{3}}{2}(\boldsymbol{D}_{1b,0}+D_{1b,1/2})\hat{z}$ | − | |

**(b)**

| Phase | Sites | $H_{\mathrm{eff}}$ | $Ho^{3+}$ moment | |
|---|---|---|---|---|
| $A1_{0°}$ | $Ho_{2a}$ | 0 | + | Fig. S2 (b) |
| | $Ho_{4b,1}$ | $\frac{3\sqrt{3}}{2}(\boldsymbol{D}_{1b,0}+D_{1b,1/2})\hat{z}$ | + | |
| | $Ho_{4b,2}$ | $-\frac{3\sqrt{3}}{2}(\boldsymbol{D}_{1b,0}+D_{1b,1/2})\hat{z}$ | − | |

**(c)**

| Phase | Sites | $H_{\mathrm{eff}}$ | $Ho^{3+}$ moment | |
|---|---|---|---|---|
| $A1_{30°}$ | $Ho_{2a}$ | $\frac{3}{2}(D_{1a,0}+D_{2a,0}+\boldsymbol{D}_{1a,1/2}+\boldsymbol{D}_{2a,1/2})\hat{z}$ | + | Fig. S2 (c) |
| | $Ho_{4b,1}$ | $\frac{3}{2}(-\boldsymbol{D}_{1b,0}+\boldsymbol{D}_{2b,0}+2D_{1b,1/2}+D_{2b,1/2})\hat{z}$ | + | |
| | $Ho_{4b,2}$ | $\frac{3}{2}(2\boldsymbol{D}_{1b,0}+\boldsymbol{D}_{2b,0}-D_{1b,1/2}+D_{2b,1/2})\hat{z}$ | + | |

## (d)

| Phase | Sites | $H_{eff}$ | $Ho^{3+}$ moment | |
|---|---|---|---|---|
| $A1_{60°}$ | $Ho_{2a}$ | $\frac{3\sqrt{3}}{2}(D_{1a,0}+D_{2a,0}+\boldsymbol{D}_{1a,1/2}+\boldsymbol{D}_{2a,1/2})\hat{z}$ | ⊕ | Fig. S2 (d) |
| | $Ho_{4b,1}$ | $\frac{3\sqrt{3}}{2}(\boldsymbol{D}_{2b,0}+D_{1b,1/2}+D_{2b,1/2})\hat{z}$ | ⊕ | |
| | $Ho_{4b,2}$ | $\frac{3\sqrt{3}}{2}(\boldsymbol{D}_{1b,0}+\boldsymbol{D}_{2b,0}+D_{2b,1/2})\hat{z}$ | ⊕ | |

## (e)

| Phase | Sites | $H_{eff}$ | $Ho^{3+}$ moment | |
|---|---|---|---|---|
| A2 | $Ho_{2a}$ | $3(D_{1a,0}+D_{2a,0}+\boldsymbol{D}_{1a,1/2}+\boldsymbol{D}_{2a,1/2})\hat{z}$ | ⊕ | Fig. S2 (e) |
| | $Ho_{4b,1}$ | $\frac{3}{2}(2\boldsymbol{D}_{1b,0}+\boldsymbol{D}_{2b,0}+2D_{1b,1/2}+D_{2b,1/2})\hat{z}$ | ⊕ | |
| | $Ho_{4b,2}$ | $\frac{3}{2}(2\boldsymbol{D}_{1b,0}+\boldsymbol{D}_{2b,0}+2D_{1b,1/2}+D_{2b,1/2})\hat{z}$ | ⊕ | |

**Table S1.** (a)-(e) The estimated effective magnetic fields in terms of the DM vectors for the consecutive five different magnetic phases. Bold letters indicate larger magnitudes of the DM vectors. The signs, +/−, denote the directions of the induced $Ho^{3+}$ moments and the values of color rings surrounding the signs indicate the strength of the moments. $Ho_{4b,1}$ and $Ho_{4b,2}$ denote two different kinds of $Ho^{3+}$ ions at the 4b sites.

**Additional data of physical properties in h-HMO**

The temperature dependence of the heat capacity divided by temperature and dielectric constant in Fig. S3(a) represents three consecutive magnetic transitions, $T_N$=75 K, $T_{SR}$ ≈37 K, and $T_{Ho}$≈5 K. The *H-T* phase diagram, constructed using various electric and magnetic properties up to 100 K and 6 T, is shown in Fig. S3(b). Under magnetic field, $T_{SR}$ shifts to lower temperature and $T_{Ho}$ tends to increase. Below $T_{Ho}$, the phase diagram develops considerable complexity with several new magnetic phases, consistent with previous work [4-7].

Figure S4(a) and (b) show 3D plots of temperature dependent $\chi''$ and $\chi'$, respectively, in various *H//c*. The *T* dependent $\chi''(\mu_0 H$=2.00 T) and $\chi'(\mu_0 H$=2.05 T) behave very similarly to the curves connecting the peak positions of *H* dependent $\chi''$ and $\chi'$, respectively in Fig. 2 (c) and (d) in the main text. No significant frequency dependence of both $\chi''(\mu_0 H)$ and $\chi'(\mu_0 H)$ was observed in the frequency dependence (*f*=0.01, 0.1, 1, 10 kHz) of $\chi''(\mu_0 H)$ and $\chi'(\mu_0 H)$ at *T*=2.2 K in the Fig. S4 (c) and (d).

The temperature dependent plot of the full-width-half-maximum (FWHM) values and the shift of $\chi'(\mu_0 H)$ between ramping up and down curves, $\Delta\mu_0 H(T)$, in Fig. 4(a) of the main text elucidate the first order nature below the temperature of the CEP. The FWHM plot was extracted from *H* derivative of *M* at various *T* (Fig. S5(a)) and hysteric behavior of $\chi'(\mu_0 H)$ at 2.0 K and 2.4 K with *H* ramping up and down in Fig. S5(b) clearly delineates the boundary between the first order transition line and the smooth crossover on either side of the CEP. We have also tried to measure a magnetic-field dependent specific heat across the CEP since this measurement is usually performed to recognize a first order phase transition. However, the strong magnetic moment of $Ho^{3+}$ at the low temperature always caused the significant rotation of the sample platform before reaching the CEP even if we minimized the amount of the sample.

The $H$-dependence of inverse $\kappa$ at $T$=2.0, 2.3, 3.0 and 5.0 K are displayed in Fig. S6. The feature of thermal conductivity measured at 2.0 K appears to be the sharpest around the critical point. Since a small amount of heat was applied during the measurement to create the temperature gradient, the average sample temperature was a bit higher than the base temperature of 2.0 K and should be close to the temperature where the critical fluctuations occur.

Figure S7(a) shows the peak intensities of the (1,0,0)* and (2,1,0)* Bragg scattering with increasing field at 2.0 K. The five different phases are well defined in the step-like features at the transitions. With the current neutron measurements, how the changes of intensities are related to the detailed $Ho^{3+}$ and $Mn^{3+}$ spin structures is not known. However, combining the results from the various experimental techniques, it is clear that the step-like features at the transitions indicate the existence of well-defined intermediate phases between zero- and high-$H$ phases. $\kappa$ at $\mu_0H$=2.05 T in Fig. S7(b) also reveals an intriguing $T$ dependence. In $\mu_0H$=0, anomalies corresponding $T_N$ and $T_{Ho}$ are clearly shown but no feature arises at $T_{SR}$. At $T_{Ho}$, an increasing behavior upon cooling in $\mu_0H$=0 changes to the linear feature in the log-log scale at $\mu_0H$=2.05 T, i.e., power-law behavior following $\sim T^{1.3}$.

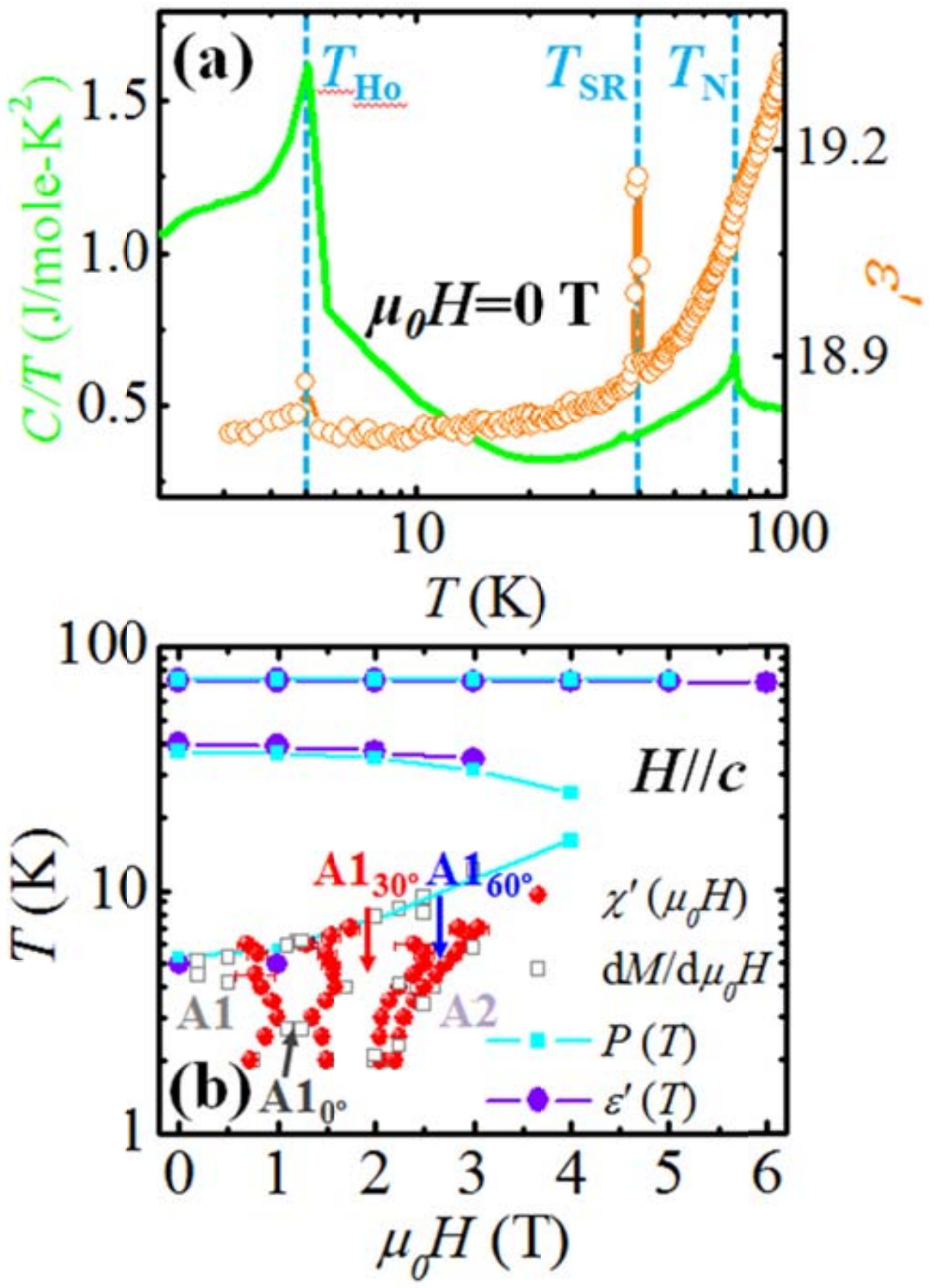

**Figure S3.** (a) $T$ dependence of the heat capacity divided by $T$ and dielectric constant representing three consecutive magnetic transitions. (b) $H$-$T$ phase diagram constructed from the real part of the AC susceptibility $\chi'(\mu_0 H)$, $H$ derivative of magnetization d$M$/d$\mu_0 H$, $T$ dependent ferroelectric polarization $P(T)$, and $T$ dependent dielectric constant $\varepsilon'(T)$, all along the $c$ axis.

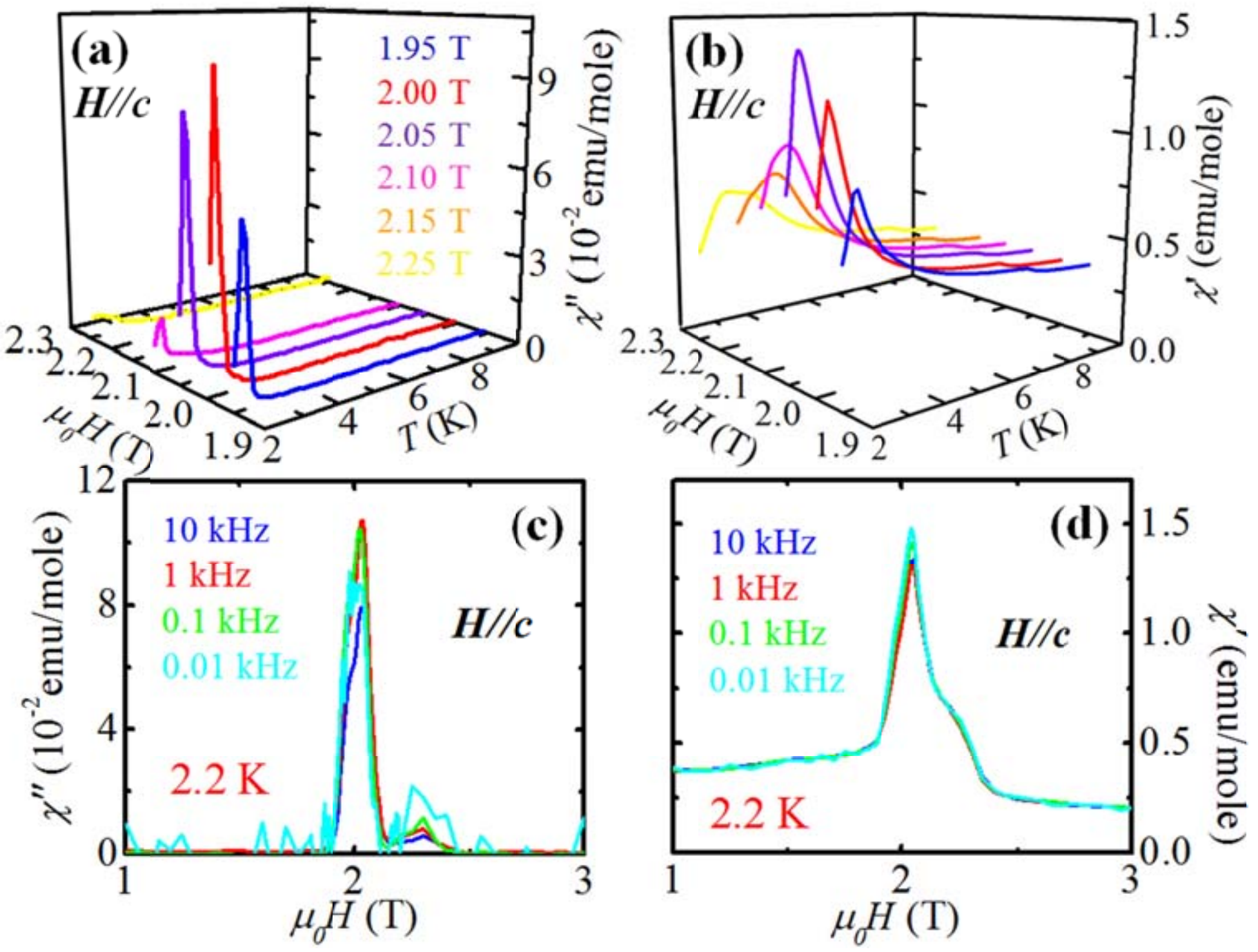

**Figure S4.** (a) and (b) 3D plots of $T$ dependent $\chi''$ and $\chi'$ , respectively, in various $H//c$. Note that 1 emu/(mol Oe) = $4\pi\times10^{-6}$ $m^3$/mol. (c) and (d) Frequency dependence ($f$=0.01, 0.1, 1, 10 kHz) of $\chi''(\mu_0 H)$ and $\chi'(\mu_0 H)$ at $T$=2.2 K.

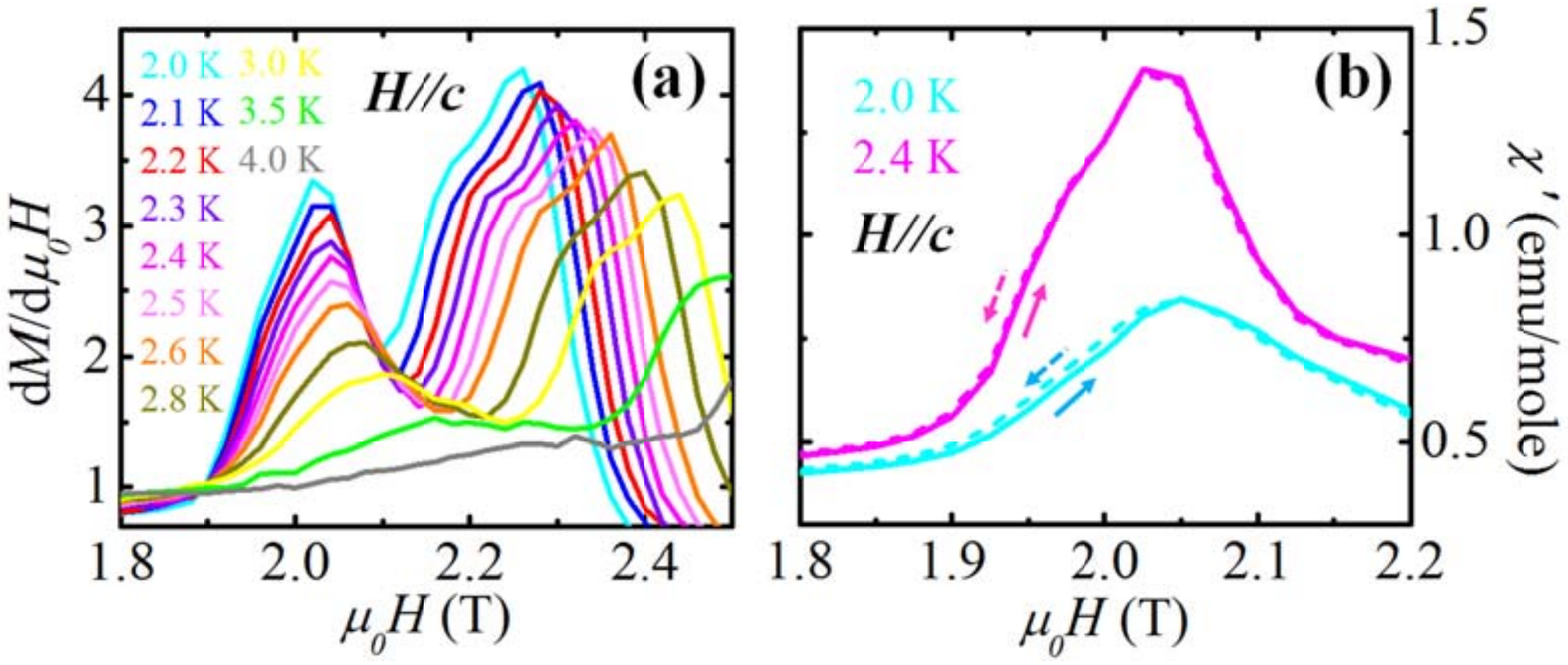

**Figure S5.** (a) and (b) $H$ derivative of $M$ at various $T$ and hysteric behavior of $\chi'$ at 2.0 K and 2.4 K with $H$ ramping up and down, respectively.

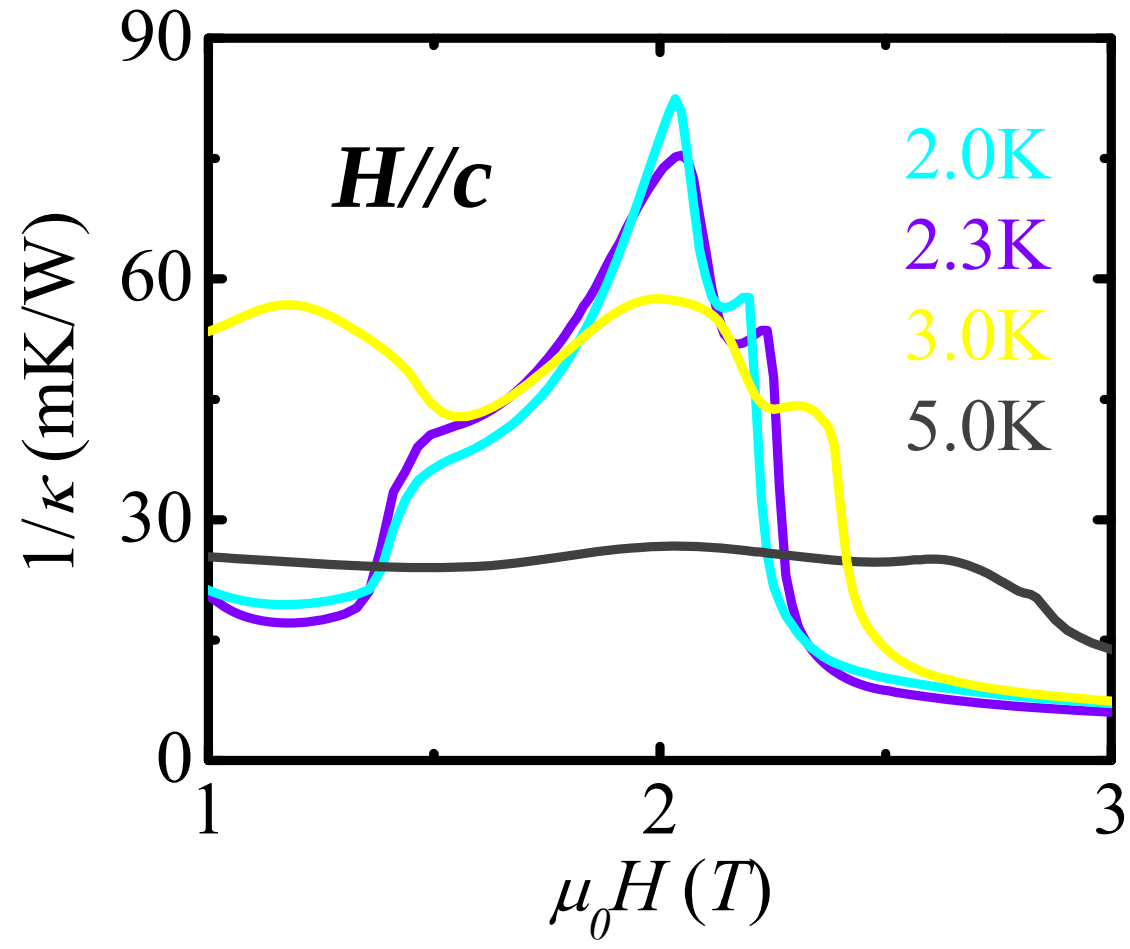

**Figure S6.** $H$-dependence of inverse $\kappa$ at $T$=2.0, 2.3, 3.0 and 5.0 K.

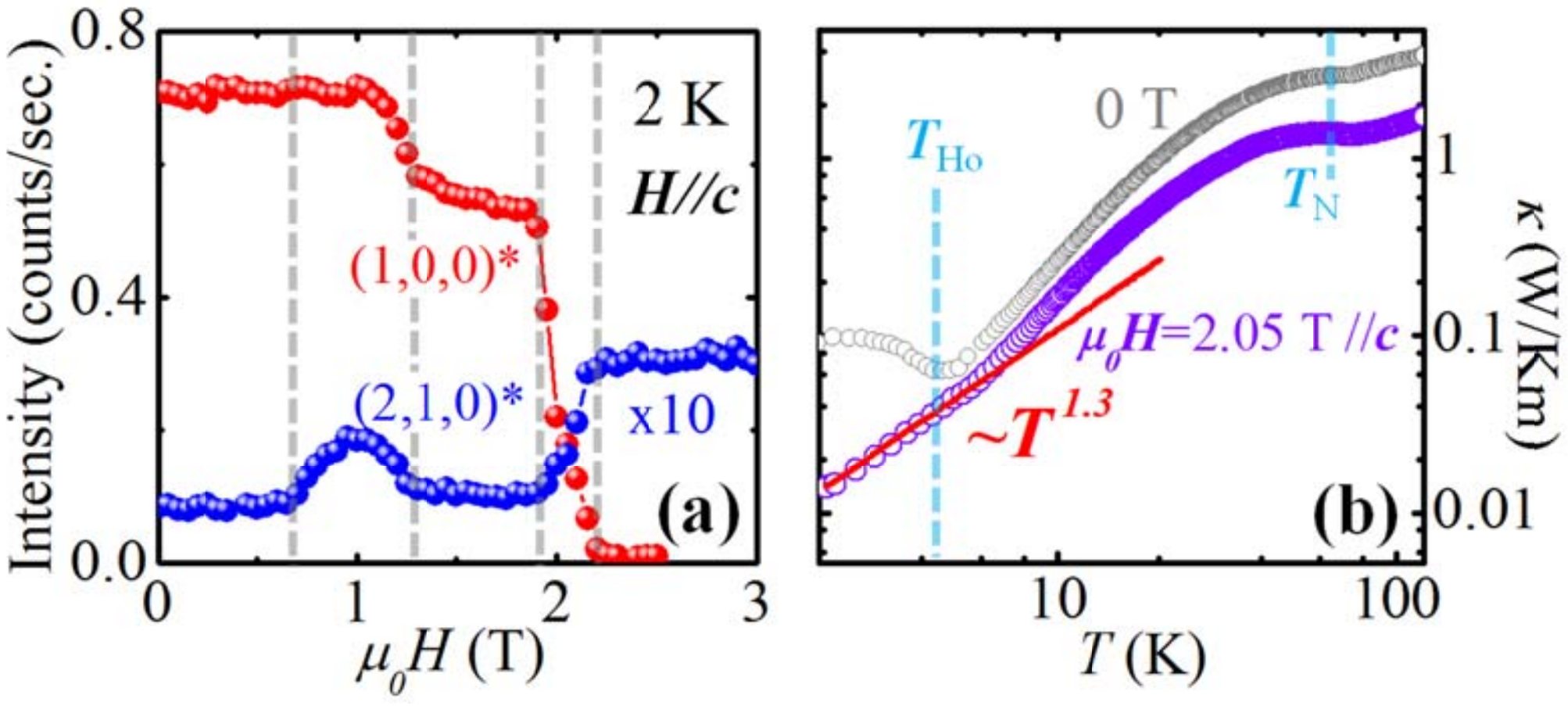

**Figure S7.** (a) $H$-dependent (1,0,0)* and (2,1,0)* magnetic Bragg peak intensities at 2.0 K with increasing magnetic field. ‘*’ denotes the orientations in the momentum space. Red and blue circles are the plots for the (1,0,0)* and (2,1,0)* Bragg reflections, respectively. Due to the weak intensity of the (2,1,0)* scattering, the data are magnified by ten times in scale. (b) Temperature dependence of thermal conductivity at $\mu_0 H$=0 and 2.05 T.

**References**

[1] T. Moriya, Phys. Rev. **120**, 91 (1960).

[2] Y. Geng *et al.*, Nano Letters **12**, 6055 (2012).

[3] S. Nandi *et al.*, Phys. Rev. Lett. **100** (2008).

[4] P. Brown, and T. Chatterji, Phys. Rev. B **77** (2008).

[5] O. P. Vajk *et al.*, J. Appl. Phys. **99** (2006).

[6] B. Lorenz *et al.*, Phys. Rev. B **71**, 014438 (2005).

[7] F. Yen *et al.*, Phys. Rev. B **71**, 180407 (2005).